\documentclass[preprint,12pt]{elsarticle}

\usepackage{graphicx}
\usepackage{amssymb}

\usepackage[utf8]{inputenc}
\usepackage{listings}
\usepackage[svgnames]{xcolor}
\usepackage{graphicx,amsmath,float}
\usepackage{latexsym,natbib,url,hyperref}
\usepackage{wrapfig,lipsum,booktabs}
\usepackage{caption,subcaption}
\usepackage{multirow,enumitem}

\newcommand{\bY}{{\bf Y}}
\newcommand{\bX}{{\bf X}}

\usepackage{bbm}

\usepackage{lineno}

\begin{document}

\begin{frontmatter}

\title{Quantifying the impact of COVID-19 on the US stock market: An analysis from multi-source information}


\author{Asim K. Dey\textsuperscript{\rm 1, 2},
 G M Toufiqul Hoque\textsuperscript{\rm 3},
Kumer P. Das\textsuperscript{\rm 4},\\
and Irina Panovska\textsuperscript{\rm 1}\\
\textsuperscript{\rm 1}University of Texas at Dallas, Richardson, TX 75080\\
\textsuperscript{\rm 2} Princeton University, Princeton, NJ 08544\\
\textsuperscript{\rm 3} Lamar University, Beaumont, TX 77705\\
\textsuperscript{\rm 4}University of Louisiana at Lafayette, Lafayette, LA 70504
}

\begin{abstract}
We develop a novel temporal complex network approach to quantify the US county level spread dynamics of COVID-19. The objective is to study the effects of the local spread dynamics, COVID-19 cases and death, and Google search activities on the US stock market. We use both conventional econometric and Machine Learning (ML) models. The results suggest that  COVID-19 cases and deaths, its local spread, and Google searches have impacts on abnormal stock prices between January 2020 to May 2020. In addition, incorporating information about local spread significantly improves the performance of forecasting models of the abnormal stock prices at longer forecasting horizons. On the other hand, although a few COVID-19 related variables, e.g., US total deaths and US new cases exhibit causal relationships on price volatility, COVID-19 cases and deaths, local spread of COVID-19, and Google search activities do not have impacts on price volatility.
\end{abstract}


\end{frontmatter}




\section{Introduction}
After the COVID-19 pandemic started spreading worldwide, the US stock market collapsed significantly  with the S\&P 500 dropping 38\% between February 24, 2020 and March 20, 2020. Similar declines have occurred in other stock indices. A number of recent studies attempt to assess the impact of the COVID-19 outbreak on the stock market. 
\cite{Nicola2020} provide a review on the socioeconomic effects of COVID-19 on individual aspects of the world economy. 
\cite{NBERw26945} analyze 
the reasons why the U.S. stock market reacted so much more adversely to COVID-19 than to previous
 pandemics that occurred in $1918-1919$, $1957-1958$ and $1968$. 
\cite{Alexander2020} gives a picture of post-COVID-19 economic world.
\cite{Enrico2020,ZAREMBA2020101597,ariascalluari2020methods}, and \cite{Cao2020} perform statistical modeling to analyze the effect of COVID-19 on the stock market price and volatility, and \cite{ludvigson_NBERw27784} study the effects of policy announcements on the stock market during the early period of the pandemic.

However, there are still a number of important questions that need to be investigated. For example,

\begin{enumerate}

\item How can we quantify the local spread dynamics of  COVID-19 and do the local spread, e.g., US county level spread of COVID-19, affect stock prices?
\item Do the number of COVID-19 cases and deaths influence stock prices?


\item Do the local spread, the number of COVID-19 cases and deaths influence the stock market's volatility? 

\item Do the Google search volumes related to COVID-19 exhibit any relationship with the stock price and volatility?

\item Do  the local spread of COVID-19, the number of cases and deaths, and Google search volumes convey any additional information about the stock price dynamics, given more conventional economic variables?

\end{enumerate}

In this study, we assess to what extent local spread affects the stock market price and volatility. In order to quantify the local spread of COVID-19, we introduce the concept of \textit{temporal network} and \textit{network motifs}. 
The network structure allows us to leverage much richer data set that includes information not only about the total number of cases, but also about the spread across counties over time.  

The stock market reacts to different local and global major events. 
\cite{Julie1996a,Worthington2004,Worthington2008TheIO,Cavallo2009}, and \cite{SHAN201236} study the impact of natural disasters, e.g., hurricanes and earthquakes, on the stock market. \cite{HUDSON2015166,Schneider2006,CHAU20141,Beaulieu2006}, and \cite{Huynh2019} evaluate the effect of political uncertainty and war on the stock market. The influences of the outbreak of infectious diseases, e.g., Ebola and SARS,  on the stock indices are assessed in \cite{Nippani2004,Alan2004,Jong-Wha2004}, and \cite{ICHEV2018153}.

Investor sentiment is another crucial determinant of stock market dynamics.
However, quantifying investor sentiment is not an easy task because of its unobservable and heterogeneous behaviors~(\cite{GARCIAPETIT201995,gao_ren_zhang_2020,Baker2007,Bandopadhyaya2005}). In recent years, due to data availability, Google search volume has become a popular index of investor sentiment~(\cite{BIJL2016150,KIM20192,Preis2013}).
\cite{BOLLEN20111} determine Twitter feeds as the moods of investors and use the \textit{Twitter mood} to predict the stock market. \cite{Alanyali2013,Schumaker2008,BOMFIM2003133}, and \cite{Albuquerque2008} evaluate the relationship between financial news and the stock market and find that news related to the asset significantly impact the corresponding stock price and volatility. \cite{ludvigson_NBERw27784} use a dynamic asset pricing model and high-frequency policy announcement news to study the effects of policy on the stock market during the COVID-19 crisis, and find that movements during the crisis have been more reflective of sentiment than substance, with the response to sentiment being a more important driver in the stock market than the responses to the actual policy actions. 

Similarly, the economics literature has shown that the inclusion of additional predictors for regional and local variables substantially improves nowcasts and forecasts for aggregate activity (\cite{HERNANDEZMURILLO2006335}), with the gains typically being concentrated during periods of large negative movements (\cite{opw_jmbc2015}). Because local spread may be indicative both of disruption in local economic activity and might be linked to local sentiment, we augment the standard aggregate models with variables that capture the local spread. 
We introduce the concept of \textit{network motifs} to model the local spread. This allows us to utilize multi-source information to quantify the impact of the spread on the US stock market. The rest of the paper is organized as follows. Section~\ref{sec:Data}  describes the data, constructs a temporal network for COVID-19 spread, and defines the variables used in the study. The methodology is described in Section~\ref{Meth}. Section~\ref{Results} presents findings and a discussion of the results. Finally, Section~\ref{sec:Conclusion} provides a conclusion.


\section{Data and Variables}\label{sec:Data}

The S\&P 500 closing price from June 3, 2019 to May 29, 2020 data are obtained from Yahoo! Finance~(\cite{Yahoo}). Google search data from January 2, 2020 to May 29, 2020 are obtained from Google Trends. We get US County level COVID-19 case data from the New York Times~(\cite{CoviddataNet}) and US county information from the US National Weather Service~(\cite{nws}).


\subsection{Abnormal stock price and volatility}
We evaluate the impact of COVID-19 on abnormalities in the S\&P 500 index. We define the daily abnormal S\&P 500 price (AP) between January 2, 2020 and May 29, 2020 by subtracting the average price of the last seven months from the daily price and by 
dividing the resultant difference from the standard deviation of the last seven months (i.e., 148 days) as follows:
\begin{eqnarray}\label{eq:AP}
AP_t= {{P_t-  {{1\over 148} \sum\limits_{i=1}^{148} P_{t-i}}} \over{\sigma_{P}}},
\end{eqnarray}
where, $P_t$ is the daily closing price for day $t$, $\sigma_{P}$ is the standard deviation of the closing price in the last 148 days~(\cite{KIM20192,BIJL2016150}). We use daily squared log returns of prices $P_t$  as a proxy for daily volatility ($Vol$)~(\cite{Brooks1998,Barndorff-Nielsen2002}): 
\begin{eqnarray}\label{eq:rt}
r_{t}=\log \left(P_t \over {P_{t-1}}\right), & Vol_t= r^2_{t}.\\\nonumber
\end{eqnarray}

\subsection{COVID-19 cases}

We study the impact of a number of COVID-19 variables ($\textbf{C}$), e.g., daily US total cases, daily US new cases,  daily World total cases, etc. on $AP_t$ and $Vol_t$. For a complete list of  COVID-19 variables see~Table~\ref{tab:OverviewA}.
We standardized each COVID-19 variable on the basis of a rolling average of the past $7$ days and corresponding standard deviation as:

\begin{eqnarray}\label{eq:CVt}
CV_t= {{C_t-  \mu_C} \over{\sigma_{C}}},
\end{eqnarray}

where, $C_t$ is a COVID-19 variable (e.g., US total cases) at day $t$,  $\mu_C$ and $\sigma_{C}$ are the mean and standard deviation of the corresponding variable within the sliding window of days $[t-k,t-1]$.

\subsection{Local Spread through complex network analysis}

A complex network  represents a  collection of elements and their inter-relationship. A network consists of a pair $G=(V,E)$ of sets, where $V$ is a set of nodes, and $E \subset V \times V$ is a set of edges, $(i,j) \in E  $ represents an edge (relationship) from node $i$ to node $j$. Here $|V|$ is the number of nodes and $|E|$ is the number of edges. The degree $d_u$ of a node $u$ is the  number of edges incident to $u$ i.e., for u,v $\in$ V and e $\in$ E, $d_u = \sum_{u \neq v} e_{u,v}$. 
A graph $G'=(V', E')$ is a \textit{subgraph} of $G$, 
if $V'\subseteq V$ and $E'\in E$.
The largest connected component (GC) is the maximal connected subgraph of $G$. 
The elements of  the $n\times n$-symmetric adjacency matrix, $A$, of $G$ can be written as
\begin{equation}
A_{ij}=\begin{cases} 1, & \mbox{if}\; (i,j)\in E \\
0, & \mbox{otherwise}.
\end{cases}
\end{equation}

\begin{figure*}[!ht]
\centering
\includegraphics[width=0.4\textwidth]{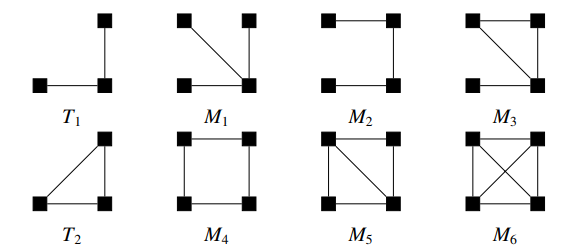}
\caption{\scriptsize All  3-node and 4-node connected network motifs.}
\label{fig:motif}
\end{figure*}

A higher-order network structure, e.g., \textit{motif}, represents local interaction pattern of the network. In a disease transmission a network motif provides significant insights about the spread of the diseases. 
For example, the presence of a dense motif or fully connected motif can increase the spread of the disease through the network, while a chain-like motif can decrease the spread of the disease~(\cite{Leitch2019}). A motif is a recurrent multi-node subgraph pattern. A detailed description of network motifs and their functionality in a complex network can be found in \cite{milo2002network,ahmed2016kais,RosasCasals_CorominasMurtra2009}, \cite{akcora2017chainlet}, and \cite{Dey2019}. Figure~\ref{fig:motif} shows all connected 3-node motifs (T) and 4-node motifs ($M$). 

\begin{figure*}[!ht]
\centering
\includegraphics[width=0.8\textwidth]{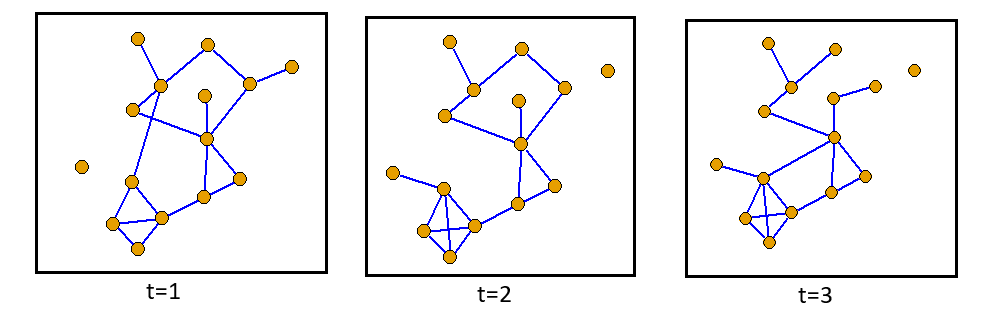}

\caption{\scriptsize A changing network shown over three time steps.}
\label{fig:Temporal_net}
\end{figure*}

Temporal Network is an emerging extension of network analysis which appears in many domains of knowledge, including epidemiology~
(\cite{Valdano2015,Demirel2017,Enright2018}), and finance~(\cite{Battiston2010,Zhao2018,Stjepan2018}). A temporal network is a network structure that changes in time. That is, a temporal network can be represented with a time indexed graph $G_t=(V(t),E(t))$, where, $V(t)$ is the set of nodes in the network at time $t$, $E(t) \subset V(t) \times V(t)$ is a set of edges in the network at time $t$. Here $t$ is either discrete or continuous. Figure~\ref{fig:Temporal_net} depicts a small $15$-node temporal network with time $t=1,2$, and $3$.

In order to quantify the county level spread of COVID-19 
we construct a complex network ($G_t$) in each day ($t$) between Jan 2, 2020 to May 29, 2020: $\mathbb{G} = \{{G}_{1}, \ldots, {G}_{\mathcal{T}}\}$, where $\mathcal{T}=130$. We evaluate the occurrences of different motifs in each $G_t$. An increase number of  motifs, i.e., $T$ and $M$, and other network features e.g., $E$, indicate a higher spread in a local community.
These increases of higher order network structures have potential impacts on $AP$ and $Vol$.


Let $C$ be the set of counties in US, $I$ is the set of COVID-19 new cases identified in $C$ on a day $t$, and $D$ is the pairwise distance matrix in miles among centroid of the counties in $C$.  We use the following three steps to construct the COVID-19 spread network ($G_t$) at time $t$ and compute the occurrences of motifs in $G_t$:
\begin{enumerate}
    \item Each County in $C$ with $\gamma$ or more COVID-19 new cases, $\gamma \in \mathbb{Z}^+$, makes a node in the network ($G_t$).
    
    \item Two counties (i.e., nodes), $i$ and $j$, are connected by an edge if (1) both counties have $\lambda$ or more COVID-19 new cases, $\lambda \in \mathbb{Z}^+$, and (2) the distance between $i$ and $j$ is less than $\delta$, $\delta \in \mathbb{R}_{\ge 0}$. Therefore, the adjacency matrix, $A^t$, is written as
    
\begin{equation}
A^t_{ij}=\begin{cases} 1, & \mbox{if}\;  I_i,I_j >\lambda~ \&~ D_{ij} > \delta \\
0, & \mbox{otherwise}.
\end{cases}
\end{equation}

 \item We compute occurrences of nodes ($V_t$), edges ($E_t$), different 3-node motif ($T(t))$,  different 4-node motifs ($M(t)$), and size of the largest connected component ($GC(t)$) in $G_t$.
\end{enumerate}

\begin{figure*}
     \centering
     \begin{subfigure}[b]{0.49\textwidth}
         \centering
         \includegraphics[width=1.\textwidth]{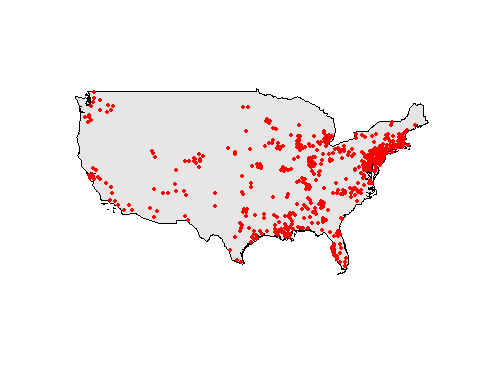}
         \caption{COVID-19 spread in US counties.}
         \label{fig:Zachary}
     \end{subfigure}
     \hfill
     \begin{subfigure}[b]{0.49\textwidth}
         \centering
         \includegraphics[width=0.90\textwidth]{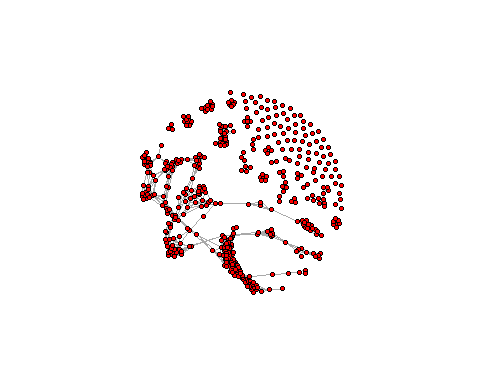}
         \caption{COVID-19 spread network.}
         \label{fig:Follower1}
     \end{subfigure}
        \caption{Local spread of COVID-19.  (a) Shows US counties with 5 or more Coronavirus cases ($\gamma=5$) on April 11, 2020. (b) Represents the corresponding spread network ($\lambda=5$, $\delta=100$)  with 514 nodes and 3831 edges.}
     	\label{fig:covid_Network}
\end{figure*}

In this study, we choose $\gamma=5$, $\lambda=5$, and $\delta=100$. That is, if two counties both have 5 or more COVID-19 cases and if the distance between these two counties is less than 100 miles they are connected by an edge. However, any appropriate choice of the parameters $\gamma$, $\lambda$, and $\delta$ can also be used to construct the COVID-19 spread network ($G_t$). For illustration,
Fig.~\ref{fig:covid_Network} shows the COVID-19 spread network in US counties on April 11, 2020. We consider different network features e.g., $E$, $T$, $M$, etc. as metrics of the local spread of COVID-19. We normalize each of the network variables based on Eq.~\ref{eq:CVt} as
\begin{eqnarray}\label{eq:GT}
SP_t= {{N_t-  \mu_N} \over{\sigma_{N}}},
\end{eqnarray}
where, $N_t$ is a network variable (e.g., $E$)  at day $t$,  $\mu_N$ and $\sigma_{N}$ are the mean and standard deviation of the corresponding variable within the sliding window of days $[t-k,t-1]$.

\subsection{Google Trend data}
A number of studies, e.g., \cite{Preis2010,BIJL2016150}, and \cite{KIM20192}, show that 
there is a significant correlation between stock variables (e.g., return,
volume, and volatility) and related Google searches, and Google search data can be used to predict future stock prices.

We investigate whether Google trend data affect the abnormal price, $AP$, and volatility, $Vol$.
We obtain the volume of the COVID-19 related daily Google searches (e.g., ``Coronavirus'') from Jan 2, 2020 to May 29, 2020. We select the location of a query in ``US'' and in the ``World''. 
We standardized each Google search variable similar to Eq.~\ref{eq:CVt} as
\begin{eqnarray}\label{eq:GT}
GT_t= {{G_t-  \mu_G} \over{\sigma_{G}}},
\end{eqnarray}

where, $G_t$ is a Google search variable at day $t$,  $\mu_G$ and $\sigma_{G}$ are the mean and standard deviation of the corresponding variable within the sliding window of days $[t-k,t-1]$. Table~\ref{tab:OverviewA} provides an overview of the data sets and variables that are used in this study. \footnote{ In a robustness check experiment, we also consider Google Trend searches for various economic policy variables, and the news based economic policy uncertainty index from \cite{bakerbloomdavis2016} as additional explanatory variables. However, while these variables were significantly correlated with abnormal returns, they did not have a causal relationship at any lag. For brevity, the results for the additional policy variables are reported in the Appendix.}  

\begin{table*}[!ht]
	\caption{Overview of the data sets.}
	
	\label{tab:OverviewA}
	\centering
	\begin{tabular}{lll} 
		\hline
			Data type  & Variables\\
	\hline
	 
	      Stock market  and  & S\&P 500 daily closing price,\\
	        Economic Uncertainty  & Economic Policy Uncertainty (EPU) Index  \\ [7pt]

			COVID-19  &   US total cases, US new cases, US total deaths, \\
		         	 &   US new death, World total cases, World new cases,  \\ 
			         &    World total deaths, World new deaths\\[7pt]

			Google Trends  &  ``Coronavirus'' US, ``Covid-19'' US,  \\
			                         &  ``Covid 19'' US, ``Covid - 19'' US, `` Coronavirus'' World,   \\
			                     
			                       &  ``Covid-19'' World,  ``Covid 19'' World, ``Covid - 19'' World \\
			                        &   ``Stimulus package''  US, ``Coronavirus stimulus''  US,   \\   & ``Stimulus''  US,  ``Stimulus check''  US,  ``irs stimulus''  US\\ [7pt]
	        Local Spread  & $V$, $E$, $GC$, $T_1$, $T_2$, $M_1$, $M_2$, $M_3$,  $M_4$, $M_5$, $M_6$ \\
\hline
		\end{tabular}

\end{table*}



\section{Methodology}\label{Meth}

We investigate the impact of COVID-19 cases and deaths, local spread of COVID-19, and COVID-19 related Google search volumes on the abnormal stock price and volatility.

\subsection{Correlation and Causality}
 A correlation test is widely used as an initial step to evaluate the relationship between the stock market and  a potential covariate~(\cite{Preis2010,Alanyali2013,Preis2013,KIM20192}). In this study, we use Spearman's rank correlation to study correlation between stock data ($AP$ and $Vol$)  and each of the COVID-19 related variables.

To assess potential predictive utilities of COVID-19 cases, local spread, and Google search interests on abnormal  price formation ($AP$) and $Vol$,  we apply the concept of Granger causality (\cite{Granger:1969}). 
The Granger causality test evaluates whether one time series is useful in forecasting another.
Let $\bY_t$, $t\in Z^+$ be a $p\times 1$-random vector ($AP_t$ or $V_t$) and let $\mathcal{F}^t_{(\bY)}=\sigma\{\bY_s: s=0,1,\ldots, t\}$ denote a $\sigma$-algebra generated from all observations of $\bY$ in the market up to time $t$. Consider a sequence of random vectors $\{\bY_t, \bX_t\}$, where $\bX$ can be either COVID-19 cases, local spread or Google search volumes. Suppose that for all $h\in Z^+$

\begin{eqnarray}
F_{t+h}\biggl(\cdot | \mathcal{F}^{t-1}_{(\bY, \bX)} \biggr)
=F_{t+h}\biggl(\cdot| \mathcal{F}^{{t-1}}_{\bY} \biggr),
\end{eqnarray}

where $F_{t+h}\biggl(\cdot| \mathcal{F}^{t-1}_{(\bY, \bX)} \biggr)$
and $F_{t+h}\biggl(\cdot| \mathcal{F}^{t-1}_{\bY} \biggr)$ are conditional distributions of $\bY_{t+h}$, given
$\bY_{t-1}, \bX_{t-1}$ and $\bY_{t-1}$, respectively. Then, $\bX_{t-1}$ is said \textit{not} to Granger cause $\bY_{t+h}$ with respect to $\mathcal{F}^{t-1}_{\bY}$. Otherwise, $\bX$ is said to Granger cause $\bY$, which can be denoted by $G_{\bX \rightarrowtail \bY}$, where $\rightarrowtail$ represents the direction of causality~(\cite{White:etal:2009,Dey2019Crudeoil,dey2020CJS}).


We fit two models where  one model includes $\bX$ and the other does not include $\bX$ (base model), and compare their predictive performance to assess causality of $\bX$ to $\bY$ using an $F$-test, under the null hypothesis of no explanatory power in $\bX$. For univariate cases we compare the following two models:

\begin{eqnarray}
\label{full_mod}
y_t=\alpha_0+\sum^d_{k=1} \alpha_ky_{t-k}+\sum^d_{k=1} \beta_kx_{t-k}+e_t,
\end{eqnarray}
versus the base model
\begin{eqnarray}
\label{red_mod}
y_t=\alpha_0+\sum^d_{k=1} \alpha_ky_{t-k}+\tilde{e}_t.
\end{eqnarray}


If $Var(e_t)$ is significantly lower than $Var(\tilde{e}_t)$, then  $x$ contains additional information that can improve forecasting of $y$, i.e., $G_{x \rightarrowtail y}$. We can also fit two linear vector autoregressive (VAR) models, with and without $\bX$, respectively, and evaluate the statistical significance of model coefficients associated with $\bX$.

\subsection{Predictive Models}

To quantify the forecasting utility of the covariates ($\bX$), i.e., COVID-19 cases, US county level spread of COVID-19, and Google searches, we develop predictive models with and without $\bX$ and compare their predictive performances. In order to conduct such a comparison, Box-Jenkins (BJ) class of parametric linear models are commonly used. However, different studies, e.g., \cite{Kane:etal:2014,dey2020CJS}, show that flexible Random Forest (RF) models often tend to outperform the BJ models in their predictive capabilities. We present the comparative analysis based on the RF models. However, any appropriate forecasting model (e.g., autoregressive integrated moving average (ARIMA($p,d,q$)),  can also be used to compare the predictive performances of the covariates.

A RF model sorts the predictor space
into a number of non-overlapping regions $R_1, R_2, \cdots, R_m$ and makes a \textit{top-down decision tree}. A common dividing technique is \textit{recursive binary splitting} process, where in each split it makes two regions $R_1=\{X|X_j < k\}$ and $R_2=\{X|X_j \ge k\}$ by considering all possible predictors $X_j$s and their corresponding cutpoint $k$ such that residual sum of squares (RSS) (Eq.~\ref{eq:lossF}) becomes the lowest.


\begin{eqnarray}\label{eq:lossF}
 RSS&=&{\sum\limits_{x_i \in R_1(j,k)} (y_i- \hat{y}_{R_1})^2} + {\sum\limits_{x_i \in R_2(j,k)} (y_i- \hat{y}_{R_2})^2}, 
\end{eqnarray}
where $\hat{y}_{R_1}$ and $\hat{y}_{R_2}$ are the mean responses for the training observations in the region $R_1(j,k)$, and in  $R_2(j,k)$, respectively.
To improve the predictive accuracy, instead of fitting a single tree, the RF technique builds a number of decision trees and averages their individual predictions~(\cite{Hastie2011}).
RF is a non-linear model (piece-wise linear). Therefore, if there is any \textit{nonlinear causality}~(\cite{Catherine2006,Emmanuel2012,Song2018}) of $\bX$ to $AP$ and $V$, a RF model 
captures this causality.

We compare the predictive performance of a baseline model (Model $P_0$), which includes only the lagged values of the abnormal price, with other proposed models which additionally include a set of covariates. The covariates are selected based on their significant correlations and causalities.  Table~\ref{table:models_ASNP} represents a description of the five models we use in our analysis.

\begin{table*}[!ht]
	\caption{Model description for abnormal price $AP$ and varying predictors.}
	\label{table:models_ASNP}	
	\centering
    
	\begin{tabular}{l|*{6}{c}r}  
		 
		\textbf{Model} &   \textbf{Predictors}      \\
			\hline	
		
		Model $P_0$  & $AP$ lag 1, $AP$ lag 2, $AP$ lag 3 \\
		
		\hline

		Model  $P_1$   & $AP$ lag 1, $AP$ lag 2, $AP$ lag 3 , \\
                & US total deaths lag 1,  US total deaths lag 2, US total deaths lag 3,\\
                & World new deaths lag 1, World new deaths lag 2, World new deaths lag 3\\

		\hline	   
		Model  $P_2$   & $AP$ lag 1, $AP$ lag 2, $AP$ lag 3 , \\
       	& Edges lag 1, Edges lag 2, Edges lag 3, GC lag 1, GC lag 2, GC lag 3,\\
       	& $T_2$ lag 1, $T_2$ lag 2, $T_2$ lag 3, $M_4$ lag 1, $M_4$ lag 2, $M_4$ lag 3 \\
       	
		\hline
			Model  $P_3$   & $AP$ lag 1, $AP$ lag 2, $AP$ lag 3,  \\
  	&  ``Covid-19'' US lag 1, ``Covid-19'' US lag 2, ``Covid 19'' US lag 1, \\
  	& ``Covid 19'' US lag 2,``Covid-19'' World lag 1, ``Covid-19'' World lag 2\\
\hline
	Model  $P_4$   & $AP$ lag 1, $AP$ lag 2, $AP$ lag 3,  \\
  	&  ``Covid-19'' US lag 1, ``Covid-19'' US lag 2, ``Covid 19'' US lag 1,\\
  	&  ``Covid 19'' US lag 2, $T_2$ lag 1, $T_2$ lag 2, \\
  	& US total deaths lag 1,  US total deaths lag 2\\
  
		\hline 
        
	\end{tabular}
\end{table*}

We consider the root mean squared error (RMSE) as measure of prediction error.
The RMSE for abnormal price modeling can be defined as
\begin{eqnarray*}
RMSE=\sqrt {{(1/n)} {\sum\limits_{t=1}^n (y_t-\hat{y_t})^2}},
\end{eqnarray*}

where $y_t$ is the test set of abnormal price ($AP$) and  $\hat{y_t}$ is the corresponding predicted value. We calculate the percentage change in prediction error (RMSE) for a specific model in Table~\ref{table:models_ASNP} with respect to model $P_0$ as

\begin{eqnarray}\label{mod}
 \Delta=\Big (1- {\Psi(P_i) \over {\Psi(P_0)} }\Big) 
 \times 100 \%,  \quad i=1,\ldots, 4,
\end{eqnarray} 

where $\Psi(P_i)$ and $\Psi(P_0)$ are the RMSE of model $P_0$ and model $P_i$, respectively.
If $\Delta > 0$, the covariate ($\bX$) is said to improve prediction of $Y$. We compare the $\Delta$ for different models, calculated for varying prediction horizons.

\subsection{Analysis of Volatility}

We evaluate the utility of COVID-19 cases and deaths, US county level spread of COVID-19, and Google searches in predicting stock market volatility. Let the conditional mean of log return of S\&P 500 price ($r_t$) be given as

\begin{eqnarray}\label{con_mean}
y_{t}= E(y_t|I_{t-1})+\epsilon_{t},
\end{eqnarray}

where  $I_{t-1}$  is the information set at time $t-1$, and $\epsilon_{t}$ is conditionally heteroskedastic error.
We build two exponential GARCH (EGARCH ($p,q$)) models, Model $0$ and Model $X$, where Model $0$ is a standard EGARCH model with no explanatory variables, and Model $X$ includes a set of explanatory variables:

\begin{eqnarray}\label{Gen_GARCH_model}\nonumber
\epsilon_{t} &=& \sigma_t \eta_t,\\ \nonumber
\text{Model $0$:}\quad \log_e {(\sigma_t^2)} &=& \omega_0+\sum^q_{i=1} \big(\omega_i \eta_{t-j} + \gamma_j \left(|\eta_{t-j}|-E |\eta_{t-j}| \right)\big)
+\sum^p_{j=1} \tau_j \log_e {(\sigma_{t-j}^2)}, \\\nonumber
\text{Model $X$:}\quad \log_e {(\sigma_t^2)} &=& \omega_0+\sum^q_{i=1} \big(\omega_i \eta_{t-j} + \gamma_j \left(|\eta_{t-j}|-E |\eta_{t-j}| \right)\big)
+\sum^p_{j=1} \tau_j \log_e {(\sigma_{t-j}^2)}+ \Lambda \bX_{t},\\
\end{eqnarray}

where $\eta_t\sim\textnormal {iid (0,1)}$, $i=1,2,\cdots,q$, $j=1,2,\cdots,p$~(\cite{Nelson1991,McAleer20104,CHANG2017,Guillaume2018,BOLLERSLEV2020411}).

We select a set of eight explanatory variables: $\bX=$ \big[US total deaths lag 1, US total deaths lag 2, $\#$ Edges lag 1, $\#$ Edges lag 2, $T_2$ lag 1, $T_2$ lag 1, ``Covid 19'' US lag 1, ``Covid 19'' US lag 2\big] with $\Lambda= \big[ \lambda_1 \quad  \lambda_2 \quad \cdots \lambda_8 \big]'$. All the explanatory variables are in the form of log returns. For simplicity we choose EGARCH (1,1) model. For EGARCH (1,1) with the assumption of $\eta_t\sim\textnormal {iid (0,1)}$  the two propose models (Eq.~\ref{Gen_GARCH_model})  reduce to
\begin{eqnarray}\label{GARCH_model}\nonumber
\text{Model $0$:}\quad \log_e {(\sigma_t^2)}= \omega_0+ \omega_i \eta_{t-j} + \gamma_j |\eta_{t-j}| +\tau_j \log_e {(\sigma_{t-j}^2)}, \\\nonumber
\text{Model $X$:}\quad \log_e {(\sigma_t^2)}= \omega_0+ \omega_i \eta_{t-j} + \gamma_j |\eta_{t-j}| +\tau_j \log_e {(\sigma_{t-j}^2)}+ \sum^8_{l=1} \lambda_l x_{lt}.\\
\end{eqnarray}

 The performances of the two models are compared based on their log likelihood, Akiake Information Criterion (AIC) and Bayesian information criterion (BIC).


\section{Results}\label{Results}

\begin{figure*}[!ht]
\centering
  \includegraphics[width=.99\textwidth,height=0.30\textheight]{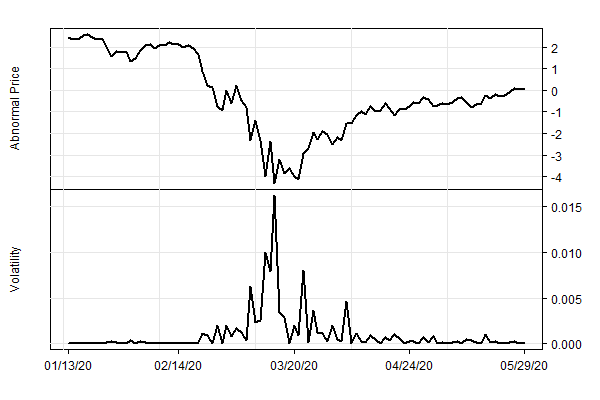}
	\caption{Time plots of abnormal price ($AP$) and volatility ($Vol$) from January 13 2020 to May 29 2020.}
	\label{fig:GR_SNP_timeplot}
\end{figure*}

We investigate the effect of COVID-19 public health crisis on the stock market, in particular, on the S\&P 500 index. We primarily focus on the impact of COVID-19 cases and deaths, local spread, and  COVID-19 related Google searches on S\&P 500. Figure~\ref{fig:GR_SNP_timeplot} shows the movements of abnormal S\&P 500 price and volatility from January 13, 2020 to May 29, 2020.
The top panel illustrates the precipitous drop of S\&P 500 price compared to the movements in prices during the last seven months (Eq.~\ref{eq:AP}). The historic high volatility (Eq.~\ref{eq:rt}) is depicted in the bottom panel.

\begin{figure*}[!ht]
     \centering
     \begin{subfigure}[b]{0.49\textwidth}
         \centering
         \includegraphics[width=.99\textwidth,height=0.30\textheight]{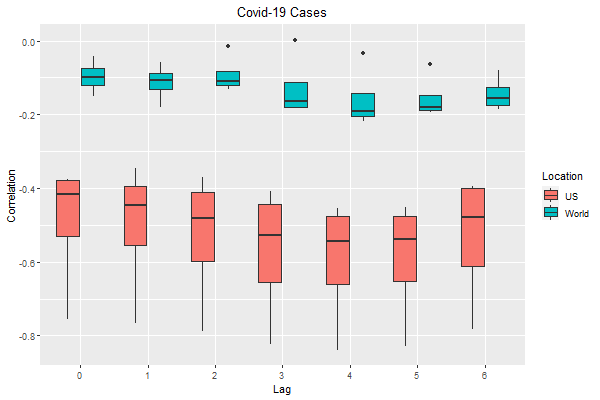}
         \caption{$AP$ and COVID-19 cases}
         \label{fig:Corr_covid19}
     \end{subfigure}
     \hfill
     \begin{subfigure}[b]{0.49\textwidth}
         \centering
         \includegraphics[width=0.99\textwidth,height=0.30\textheight]{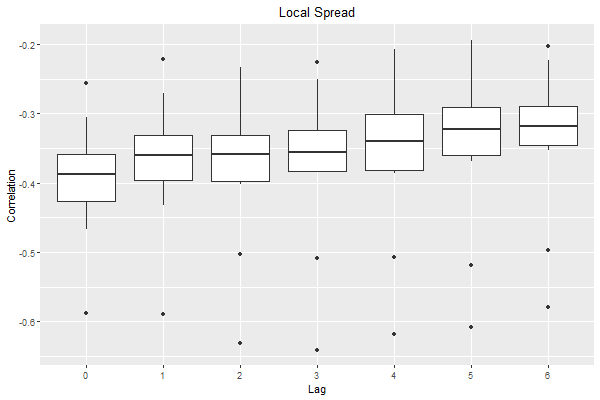}
         \caption{$AP$ and COVID-19 local spread.}
         \label{fig:Corr_spread}
     \end{subfigure}
        	\caption{(Spearman) Correlations between Covid-19 and abnormal S\&P 500. Correlations of eight COVID-19 variables in each lags are summarized in a box plot.}
        \label{fig:Correlation}
\end{figure*}

We start our analysis with the Spearman's rank correlation test. We calculate correlations between the daily abnormal  S\&P 500 closing price, $AP$ and the daily COVID-19 cases and deaths, and daily occurrences of higher order structures in the spread network at different time lags. For example, at $lag~1$ we compute correlation of $AP$ at day $t$ with COVID-19 cases and deaths, and higher order network structures, all at day $t-1$. These lag correlations evaluate the directionality of the relationships. 
Figure~\ref{fig:Corr_covid19} shows the box plots which combine  correlations between each COVID-19 cases and deaths variable and $AP$ at different $lag$. Here we build two box plots at each lag: one for COVID-19 cases and deaths in the US (four variables), and another for COVID-19 cases and deaths in the World (four variables). Similarly, Figure~\ref{fig:Corr_spread} represents the box plots that combined  correlations between each eleven local spread variable and $AP$ at different $lag$. 


We find that there exists significant (negative) correlation between COVID-19 cases and deaths in US and abnormal S\&P 500 in all six lags, $lag=1,2, \cdots, 6$. However, there is no significant correlation between COVID-19 cases and deaths in the entire world and abnormal S\&P 500 ($p$-value $>$ 0.05) in any lag (see Table~\ref{table:Corr_Covid19} in Appendix). We also find that all the local spread variables are  significantly (negative) correlated ($p$-value $<$ 0.05) with  abnormal S\&P 500 in every $lag=1, 2, \cdots, 6$.
That is, US county level spread of COVID-19 adversely affects the price of S\&P 500.  However, it is anticipated that the strength of correlations of local spread variables will gradually decrease in higher lags, which is also reflected in Figure~\ref{fig:Corr_spread}. Some of the COVID-19 related Google searches, e.g., ``Covid-19'' in US and  ``Corona'' in world are also significantly correlated ($p$-value $>$ 0.1) with abnormal S\&P 500 in different lags (Table~\ref{table:Corr_GT} in Appendix).

We also investigate the potential impact of COVID-19 cases and deaths, its local spread, and related Google searches on S\&P 500 price formation and risk, i.e., volatility. 
Table~\ref{table:causality_Covid} and Table~\ref{table:causality_Spread}
present summaries of the Granger causality tests for predictive utility of  COVID-19 cases and deaths, and county level local spread, respectively. Here the direction of causality is denoted by $\rightarrowtail$. 

We find that US total new cases and US total death have significant predictive impacts on price and volatility. US total number of cases have predictive relationship only with volatility in few lags. Among world COVID-19 cases and deaths only total new deaths have causality on price and volatility.
Almost all the local spread variables have predictive impact on price, but none of them except $\#$ Edges at $lag~1$ have causality on volatility. That is, county level spread of COVID-19 significantly influence abnormal price formation, but, surprisingly, they do not have causal linkage with the volatility. Table~\ref{table:causality_GT} in Appendix shows that a number of Google search variables have causality effects on abnormal price. However, only ``Coronavirus'' US and ``Covid 19'' US have predictive impacts on volatility at very few lags.
The price level responds to local spread and to sentiment, whereas the volatility is mostly affected by national factors.

\begin{table*}[!ht]
\small
\caption{Summary of G-causality analysis of COVID-19 cases and deaths on abnormal S\&P 500 ($y$) on different lag effects (day). $P$ and $Vol$ denote significance in price and volatility, respectively.  Blank space implies no significance. Confidence level is 90$\%$.}
	\label{table:causality_Covid}
	\centering
	\begin{tabular}{c | *{7}{c} r}   
    	       &      &    &   & \textbf{Lag} &  & & \\
		\textbf{Causality}~~~~~~~~   &  1 &     2  &  3   &  4  &  5 &  6  &  7  \\
		
		\hline
    US total cases $\rightarrowtail$ $y$          & - & - & - & $Vol$&  - &  $Vol$& -\\  	
    US total deaths $\rightarrowtail$ $y$     & P/$Vol$&  $Vol$& P/$Vol$& P/$Vol$& P/$Vol$&  P/$Vol$& P/$Vol$\\ 
    
    US new cases $\rightarrowtail$ $y$     & P/$Vol$& - &   - & $Vol$& P & P/$Vol$& P/$Vol$\\
     US new deaths $\rightarrowtail$ $y$      & -  &  - & - & - & - & -& -\\

	 \hline	
    
     World total cases $\rightarrowtail$ $y$    &  -   & -  & - &  - &  - & - & -\\ 
	 World total deaths $\rightarrowtail$ $y$   & -   & - & - & - & - & - & -\\  
	World new cases $\rightarrowtail$ $y$     & -   & -  & - & - & - & - & -\\ 
	World new deaths $\rightarrowtail$ $y$   & -  &  P  & P & P & $Vol$& $Vol$&  -\\

       \hline	
                     
\end{tabular}
\end{table*}

\begin{table*}[!ht]
\small
\caption{Summary of G-causality analysis of COVID-19 spread on abnormal S\&P 500 ($y$) on different lag effects (day). $P$ and $Vol$ denote significance in price and volatility, respectively.  Blank space implies no significance. Confidence level is 90$\%$.}
	\label{table:causality_Spread}
	\centering
	\begin{tabular}{c | *{7}{c} r}   
    	       &      &    &   & \textbf{Lag} &  & & \\
		\textbf{Causality}~~~~~~~~   &  1 &     2  &  3   &  4  &  5 &  6  &  7  \\
		
		\hline
   $\#$ Edges $\rightarrowtail$ $y$ & $Vol$ &  P & P &  P  &  P &  P &  P\\ 
	 GC $\rightarrowtail$ $y$        & P &  P &  P & P  &  P & P &   P \\  
	$T_1$ $\rightarrowtail$ $y$      & P &  P &  - & -&  - & -& -\\ 
	$T_2$ $\rightarrowtail$ $y$      & P  &   P &  P & P &  P &  P &   P \\ 
	 \hline	
	 
    $V_1$ $\rightarrowtail$ $y$         & - & -  &   P & - & - &  - & -\\  	
    $V_2$ $\rightarrowtail$ $y$         &  P &   P &  - &   P &   P  & - &   P\\ 
   $V_3$ $\rightarrowtail$ $y$       &  - &  - &    - &  - & - &  - &  -\\
    $V_4$ $\rightarrowtail$ $y$      & -&    P &    P &   P &   P &  P &  P  \\
   
     $V_5$ $\rightarrowtail$ $y$      &  - &   P &    P &   P  &    P &  P &  P \\
    $V_6$ $\rightarrowtail$ $y$      & - &  - &  - &  P & - & - & - \\
 Total  $\#$ $V$ $\rightarrowtail$ $y$    & - & - & - &  P & - & - & - \\

             		 \hline	
                     
\end{tabular}
\end{table*}

Now we turn our analysis to compare the predictive performance of models described in Table~\ref{table:models_ASNP}. Table~\ref{table:RMSE} percents prediction errors based on Eq.~\ref{mod} calculated for varying prediction horizons $h=1, 2, \ldots, 6$. 
For short term forecasting horizons ($h=1, 2$, and $3$) model $P_3$, which is based on Google search variables yields more accurate performance. 
For longer term forecasting horizons ($h=4, 5$, and $6$), model $P_2$ containing information from local spread delivers the most competitive results, followed by model $P_4$, which contains information from COVID-19 deaths, local spread, and Google searches.

\begin{figure}[!ht]
\subcaptionbox{h=1 day.\label{fig:Obs_Pred_h12}}{\includegraphics[width=0.48\textwidth]{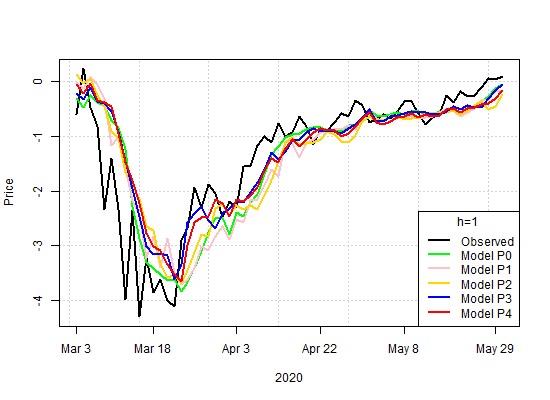}}	
	\subcaptionbox{h=2 days.  \label{fig:Obs_Pred_h7_Bitcoin_V2}}{\includegraphics[width=0.48\textwidth]{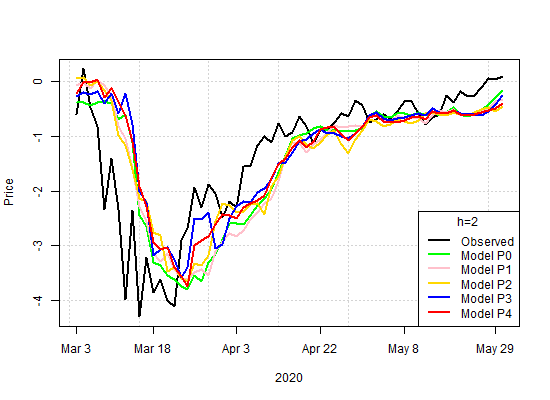}}
 \caption {Abnormal price prediction for March 2020 to May 2020 with 1, and 2 day horizons.} 
\label{fig:h_allpred}
\end{figure}

Figure~\ref{fig:h_allpred} represents a comparison of the observed data with fitted values from baseline model (model~$P_0$) and four other models, i.e., model~$P_1$, ~$P_2$, ~$P_3$, and~$P_4$. For  1 day horizon model~$P_3$ yield a noticeably higher predictive accuracy followed by model~$P_4$. For 2 day horizon, although it is expected that the prediction performances of all models deteriorates compare to their performances for 1 day horizon, model~$P_3$ again delivers the best prediction accuracy.

\begin{table*}[!ht]
 \caption{ Predictive utilities ($\Delta$) of models in Table~\ref{table:models_ASNP} over the baseline model (Model~$P_0$) for different prediction horizons.}
	\label{table:RMSE}
	\centering
    	\begin{tabular}{c *{5}{c} r}   
        \hline
    h   &  Model $P_1$   &   Model $P_2$    & Model $P_3$  &  Model $P_4$ \\
       \hline
    1  & -0.411 & -5.305 & 7.219 & 2.086\\	
    2  & 0.171  & -1.257 & 2.279  & 0.042 \\
    3 &  1.549 &   3.242 &   3.477 &  2.397\\
   4 &   1.463 &  3.579 &  3.368 & 2.672\\
    5 & 1.410  &  3.843 &  2.718 &  2.922\\
    6 &  0.962 &   3.898 &   2.718 &  3.107\\
	 \hline	
	\end{tabular}
\end{table*}


We now evaluate the influence of  COVID-19 cases and deaths, US county level spread of COVID-19, and Google searches on S\&P 500 volatility. 
A comparison of the two EGARCH models, \textit{Model 0} and \textit{Model X} (Eq.~\ref{GARCH_model}), including the estimated parameters of the explanatory variables for \textit{Model X} are presented in Table~\ref{table:EGARCH}.
All EGARCH coefficients except the constant term ($\omega_0$) are statistically significant in both models. However, the coefficients estimates of all the covariates in \textit{Model X}  are not statistically significant.

\begin{figure}[!ht]
\centering
  \includegraphics[width=0.65\textwidth]{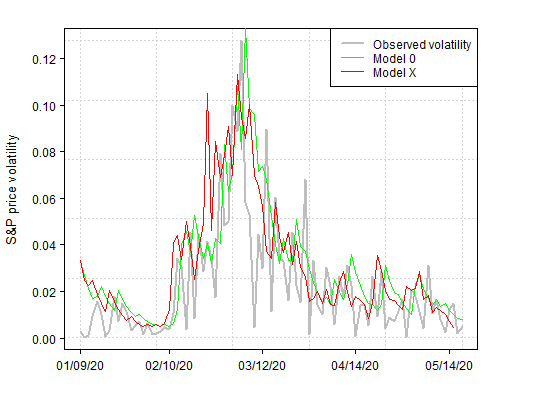}
	\caption{Time plots of $AP$ and $Vol$ from January 13 2020 to May 29 2020.}
	\label{fig:volatility_com} 
\end{figure}

We also examine the goodness of fit of the two models by comparing their log likelihood, Akiake Information Criterion (AIC) and Bayesian information criterion (BIC). We find that \textit{Model 0} tends to describe the S\&P 500 volatility more accurately than the volatility model with covariates, \textit{Model X}. That is, COVID-19 cases and deaths, its local spread and Google searches do not significantly influence the S\&P 500 volatility. Figure~\ref{fig:volatility_com} also suggests that \textit{Model 0} captures the spikes of the price returns more accurately than \textit{Model X}.

\begin{table*} [!ht]
	\caption{Estimates of EGARCH models for S\&P 500 price volatility. $^{***}~p<0.01$, ~$^{**}~p<0.05$,~$^{*}~p<0.1$.}
	\label{table:EGARCH}
	\begin{center}
	\begin{tabular}{l*{6}{c}r} \hline

			 &  Model $X$  &   &  Model $0$  & \\ 
			Parameter  & Coef. & $t$ value   &  Coef. & $t$ value   \\
			\hline
			$\omega_0$ &   -0.611  & -1.287 &   -0.392 &  -1.394\\
			$\omega$  & -0.517 & -3.249$^{***}$  & -0.484  & -3.443$^{***}$ \\
	      	$\gamma$ &  0.514 & 2.387$^{***}$ &  0.513  &  2.871$^{***}$\\
	      		$\tau$ &  0.937 & 16.192$^{***}$  & 0.955 & 2.871$^{***}$\\
		US total deaths lag 1 ($\lambda_1$)  & -0.867 & -0.864 & &   \\
		US total deaths lag 2 ($\lambda_2$) &   -0.558 &-0.609 & &    \\
		$\#$ Edges lag 1 ($\lambda_3$) & -0.260 & -1.235 & &   \\
		$\#$ Edges lag 2 ($\lambda_4$) &   0.050 & 0.221 & &  \\
	$T_2$ lag 1 ($\lambda_5$)  & -0.514 & -0.939 & &    \\
	$T_2$ lag 1 ($\lambda_6$) &   -0.067 & -0.120 & &   \\
	``Covid 19'' US lag 1 ($\lambda_7$) & -0.132 & -0.167 & &   \\
	``Covid 19'' US lag 2 ($\lambda_8$) &   -0.476 & -0.621  & &  \\
		 & &   & &  \\
		
	 Log-likelihood & 214.049	 & & 218.258 & \\
	 AIC &-4.540 & & 	-4.709 & \\
	 BIC & -4.205	& & -4.599 & \\
				\hline
	
		\end{tabular}
	\end{center}
\end{table*}

COVID-19 related factors do not seem to play a role when it comes to explaining the dynamics of the volatility. One potential explanation for this is that the movements in the volatility are driven by national-level sentiment about policy or about policy uncertainty. As a robustness check, we perform an experiment where we explore the correlations and Granger causality between Google trend searches for macroeconomic policy variables. Because the data for the COVID-19 spread is available at daily frequency, we use daily data in all of our specifications. Therefore, we proxy for investor sentiment about policy by using Google Trend searches and the policy uncertainty index from \cite{bakerbloomdavis2016} rather than higher frequency 30 minute windows around policy announcements as in \cite{ludvigson_NBERw27784}. The results are reported in the Appendix. Our results are very similar to the results for the benchmark specifications with the volatility being affected by national factors. While sentiment about policy is correlated with abnormal prices, we only find Granger causality in several cases related to fiscal policy searches, and only at higher lags. On the other hand, sentiment about policy Granger-causes volatility at all lags.

\section{Conclusion} \label{sec:Conclusion}

The aim of this paper is to evaluate whether  COVID-19 cases and deaths, local spread of COVID-19, and Google search activity 
explain and predict US stock market plunge in the spring of 2020.
 We develop a modeling framework that systematically evaluates the correlation - causality - predictive utility of each of the COVID-19 related features on stock decline and stock volatility. In order to quantify local spread of COVID-19 we construct a temporal spread network and study the dynamics of higher order network structures as a measure of local spread.
We find that COVID-19 cases and deaths, its local spread, and Google search activities related to COVID-19 have contemporary relationships and predictive abilities on abnormal stock prices. Our results indicate that COVID-19 cases and deaths, and its local spread not only unprecedentedly disrupt economic activity and cause a collapse in demand for different goods but also they make investors panic and increase their anxiety. The anxiety is also reflected in Google search intensity for COVID-19. These shocks affect investment decisions and the subsequent stock price dynamics. On the other hand,  very few COVID-19 variables have causal relationship on volatility. However, standard EGARCH models for the volatility show that COVID-19 cases and deaths, its local \textit{spread}, and Google search volumes do not have impact on volatility. Different forms for the volatility measure~\cite{MOLNAR201220,KIM20192,BIJL2016150} lead to the same conclusions.

Overall, the volatility is mostly affected by national factors and incorporating higher-order information about local spread does not significantly improve the forecasting performance of the models. However, the local spread are significantly linked to abnormal price returns. Furthermore, incorporating information about the local spread significantly improves the predictive performance of the models for the abnormal price level.

\section{Appendix}

\begin{table*}[!ht]
\small
\caption{Spearman correlations between COVID-19 cases and abnormal S\&P 500. blue color indicates significant correlation ($p$-values $<$ 0.05), while black color represents non-significant correlation ($p$-values $>$ 0.05).}
	\label{table:Corr_Covid19}
	\centering
	\begin{tabular}{c | *{7}{c} r}   
     &      &    &   & \textbf{Lag} &  & & \\
       
	 & 0 &  1 &     2  &  3   &  4  &  5 &  6   \\
\hline
	US total cases         & {\color{blue}-0.45}&{\color{blue}-0.48}& {\color{blue}-0.53}&{\color{blue}-0.59}&{\color{blue}-0.60}&{\color{blue}-0.59}&{\color{blue}-0.55}&\\ 

     \hline
    US total deaths    &{\color{blue}-0.75}&{\color{blue}-0.76}&{\color{blue}-0.78}&{\color{blue}-0.82}& {\color{blue}-0.83}&{\color{blue}-0.82}&{\color{blue}-0.78}&\\
    
    \hline
    US new cases    &{\color{blue}-0.37}& {\color{blue}-0.41}&{\color{blue}-0.42}&{\color{blue}-0.45}&{\color{blue}-0.48}&{\color{blue}-0.48}&{\color{blue}-0.40}&\\
    
    \hline
     US new deaths  &{\color{blue}-0.37}&{\color{blue}-0.34}&{\color{blue}-0.36}&{\color{blue}-0.40}&{\color{blue}-0.45}&{\color{blue}-0.45}&{\color{blue}-0.39}&\\

		\hline
     World total cases  & -0.04   & -0.05&-0.01 &0.00&-0.03&-0.06&-0.07&\\ 
       
     	 \hline

         World total deaths   & -0.08&-0.11&-0.11&-0.18&-0.20&-0.19& -0.17&\\ 
       
     	 \hline	
   
	World new cases        &-0.14&-0.18&-0.13&-0.15&-0.17&-0.18&-0.18&\\
	
		 \hline	
	World new deaths        &-0.11& -0.09&-0.10&-0.17&-0.21&-0.17&-0.14&\\

	 \hline

\end{tabular}
\end{table*}

\begin{table*}[!ht]
\small
\caption{Spearman correlations between Local spread variables and abnormal S\&P 500. blue color indicates significant correlation ($p$-values $<$ 0.05). A non-significant correlation ($p$-values $>$ 0.05) is presented by black color.}
	\label{table:Corr_Net}
	\centering
	\begin{tabular}{c | *{7}{c} r}   
       	       &      &    &   & \textbf{Lag} &  & & \\
       	  
		  & 0&  1 &     2  &  3   &  4  &  5 &  6   \\
		
		\hline
   Edge  &{\color{blue}0.60}&{\color{blue}-0.58}&{\color{blue}-0.63}&{\color{blue}-0.64}&{\color{blue}-0.61}&{\color{blue}-0.60} &{\color{blue}-0.57}&\\
     
     	 \hline

       GC   &{\color{blue}0.61} & {\color{blue}-0.43} &{\color{blue}-0.50}&{\color{blue}-0.50}&{\color{blue}-0.50}&{\color{blue}-0.51}&{\color{blue}-0.49}\\
         
     	 \hline	
   
	$T_1$      &{\color{blue}0.95}& {\color{blue}-0.35}&{\color{blue}-0.35}&{\color{blue}-0.34}&{\color{blue}-0.31}&{\color{blue}-0.30}&{\color{blue}-0.30}\\
	
		 \hline	
    $T_2$        &{\color{blue}0.90}& {\color{blue}-0.35}&{\color{blue}-0.36} &{\color{blue}-0.35}&{\color{blue}-0.34}&{\color{blue}-0.32}&{\color{blue}-0.32}&
    \\

	 \hline	

      $M_1$        & {\color{blue}0.82}& {\color{blue}-0.22}&{\color{blue}-0.23} &{\color{blue}-0.22}&{\color{blue}-0.20}&-0.19&-0.20&
      \\
     \hline
    $M_2$  &{\color{blue}0.93}& {\color{blue}-0.31}&{\color{blue}-0.30}& {\color{blue}-0.30}&{\color{blue}-0.28}&{\color{blue}-0.27}&{\color{blue}-0.27}&

    \\
   
    \hline
    $M_3$    &{\color{blue}0.98}& {\color{blue}-0.35}&{\color{blue}-0.35}&{\color{blue}-0.35}&{\color{blue}-0.32}&{\color{blue}-0.31}&{\color{blue}-0.30}&\\
    
    \hline
     $M_4$ &{\color{blue}0.84}&{\color{blue}-0.27}&{\color{blue}-0.26} &{\color{blue}-0.25}&{\color{blue}-0.25}&{\color{blue}-0.22}&{\color{blue}-0.22} &
     \\
     \hline
     $M_5$ & 0.84 & {\color{blue}-0.36}&{\color{blue}-0.35}&{\color{blue}-0.35}&-{\color{blue}0.33} &{\color{blue}-0.32}&{\color{blue}-0.31}&
     \\
    \hline
     
     $M_6$   &{\color{blue}0.91}& {\color{blue}-0.40}&{\color{blue}-0.40}&{\color{blue}-0.38}&{\color{blue}-0.38}&{\color{blue}-0.36}&{\color{blue}-0.35}&\\
     
    \hline
   $Tot_M$ &{\color{blue}0.93}&{\color{blue}-0.39}&{\color{blue}-0.39}&{\color{blue}-0.38}&{\color{blue}-0.37}&{\color{blue}-0.35}&{\color{blue}-0.33}&\\
     
     \hline
\end{tabular}
\end{table*}



\begin{table*}[!ht]
\small
\caption{Spearman correlations between google trend and abnormal S\&P 500. A significant correlation ($p$-values $<$ 0.05) is represented by blue color, while black color indicates a non-significant correlation ($p$-values $>$ 0.05).}
	\label{table:Corr_GT}
	\centering
	\begin{tabular}{c | *{7}{c} r}   
       	       &      &    &   & \textbf{Lag} &  & & \\
       	  
	  & 0&  1 &     2  &  3   &  4  &  5 &  6   \\
		
		\hline
    ``Coronavirus'' US     & {\color{blue}0.02}  & -0.03 &-0.10 &-0.11  & -0.08 & -0.07 & -0.10  &  \\ 
      
     	 \hline	
     ``Corona'' US     & 0.25  & 0.18 & 0.12 & 0.11 & 0.09 & 0.05 & 0.02  &  \\ 
      
     	 \hline

      	``Covid-19'' US       &{\color{blue}-.26} & {\color{blue}-0.31} & {\color{blue}-.28} & {\color{blue}-0.28} & {\color{blue}-0.29} & {\color{blue}-0.29} & {\color{blue}-0.32} & \\

     	 \hline


	 \hline	

``Covid 19" US &{\color{blue}-0.09}   &-0.15 & -0.17&{\color{blue}-0.21}  &{\color{blue}-0.25} &{\color{blue}-0.29}&{\color{blue}-0.30}&\\ 
     
     \hline
     
``Coronavirus" World &{\color{blue}0.10}   & 0.08 &.01 &-0.03 &-0.05 & -0.08&-0.06&\\
    \hline
 ``Corona'' World     & {\color{blue} 0.26}  & {\color{blue} 0.22}  & 0.16& 0.11 & 0.06 & 0.02  &  0.00 &  \\    
  
           		 \hline	
     ``Covid-19" World   &{\color{blue}-0.04} &-.07 &-0.08 &-0.10&-0.11&-0.12& -0.14 &\\

    \hline
    ``Covid 19" World   & -0.11 &-0.11&-0.12  &-0.15 &-0.15  &-0.21& -0.25 &\\


             		 \hline	
\end{tabular}
\end{table*}

\begin{table*}[!ht]
\small
\caption{G-causality analysis of Google searches on abnormal S\&P ($y$) on different lag effects (day). $P$ and $Vol$ denote significance in price and volatility, respectively.  Blank space implies no significance. Confidence level is 90$\%$.}
	\label{table:causality_GT}
	\centering
	\begin{tabular}{c | *{7}{c} r}   
    	           &      &    &   & \textbf{Lag} &  & & \\
		\textbf{Causality}~~~~~~~~   &  1 &     2  &  3   &  4  &  5 &  6  &  7  \\
		
		\hline
     ``Coronavirus'' US $\rightarrowtail$ $y$    & -  & -  &  - &   -  & -  &  - & -\\
	``Covid-19'' US $\rightarrowtail$ $y$        & P & P  & P  &  P & P & P & P \\
	``Covid 19'' US $\rightarrowtail$ $y$         &  $Vol$ & P & P & P & P & - &  - \\ 
	``Covid - 19'' US $\rightarrowtail$ $y$       & -  & P  & - &  - & - & - & -\\

	 \hline	
    ``Coronavirus'' World $\rightarrowtail$ $y$     & -        & - & P  &  - &  - & - & -\\  
      ``Covid-19'' World $\rightarrowtail$ $y$       & P       & P  & P & P & P & P & P\\
        ``Covid 19''  World $\rightarrowtail$ $y$   &  -      & -  & - &  - &  - &  - & -\\  
     ``Covid - 19'' World $\rightarrowtail$ $y$     & P          & P &  P &  P & - &  - &  -\\

	 \hline

\end{tabular}
\end{table*}


\begin{table*}[!h]
\small
\caption{Spearman correlations between  Economic Policy Uncertainty (EPU) Index and  
google trend in US related to economic policy, and abnormal S\&P 500. A significant correlation ($p$-values $<$ 0.05) is represented by blue color, while black color indicates a non-significant correlation ($p$-values $>$ 0.05).}
	\label{table:Corr_GT}
	\centering
	\begin{tabular}{c | *{7}{c} r}   
       	       &      &    &   & \textbf{Lag} &  & & \\
       	  
	  & 0 &  1 &     2  &  3   &  4  &  5   \\
				\hline
    EPU    & {\color{blue}-0.276}  &  {\color{blue}-0.277} & {\color{blue}-0.300}  &   {\color{blue}-0.280} &   {\color{blue}-0.268} & {\color{blue}-0.308}\\

		\hline
    ``Unemployment benefit''      & {\color{blue}-0.382}  &  {\color{blue}-0.388} & {\color{blue}-0.357}  &   {\color{blue}-0.340} &   {\color{blue}-0.285} & {\color{blue}-0.242}\\ 
      
     	 \hline	
      ``Stimulus package''     & {\color{blue} -0.177} &  -0.150 & -0.137 & -0.151 & -0.164 & -0.148  \\ 
      
     	 \hline	
     
      	 ``Coronavirus stimulus''    &  {\color{blue} -0.192}    &{\color{blue} -0.194 } & {\color{blue}-0.189} & -0.244  & -0.249  & {\color{blue}-0.223} \\

	 \hline	

  ``Stimulus'' & -0.159   & -0.155 & -0.171  & -0.162 &  -0.171  &  -0.162\\ 
     
     \hline
     
  ``Stimulus check'' &  -0.058     &  -0.0374 &  -0.073 & -0.058  & -0.071 &  -0.057 \\
    \hline
   ``irs stimulus''     & {\color{blue} -0.289 }  & {\color{blue} -0.282 }  &  {\color{blue} -0.282} & {\color{blue} -0.323} & {\color{blue}-0.314} &  {\color{blue}-0.305}\\    
  
           		 \hline	
\end{tabular}
\end{table*}


\begin{table*}[!ht]
\small
\caption{G-causality analysis of  Economic Policy Uncertainty (EPU) Index  and  
google trend in US related to economic policy on abnormal S\&P ($y$) on different lag effects (day). $P$ and $Vol$ denote significance in price and volatility, respectively.  Blank space implies no significance. Confidence level is 90$\%$.}
	\label{table:Corr_GT}
	\centering
	\begin{tabular}{c | *{7}{c} r}   
       	       &      &    &   & \textbf{Lag} &  & & \\
       	  
	  &  1 &     2  &  3   &  4  &  5  & 6 & 7 \\
				\hline
    EPU  $\rightarrowtail$ $y$    & - &  -& -  &  - &  - & -&  - \\

		\hline
    ``Unemployment benefit''  $\rightarrowtail$ $y$     & - &  -& -  &  - &  - & -&  - \\ 
      
     	 \hline	
      ``Stimulus package''  $\rightarrowtail$ $y$      & $Vol$ & $Vol$& $Vol$ & $Vol$ &  $Vol$ & $Vol$ & $Vol$ \\ 
      
     	 \hline	
     
      	  ``Coronavirus stimulus''    $\rightarrowtail$ $y$      & $Vol$ & $Vol$& $Vol$ & P/$Vol$ &  P/$Vol$ & P/$Vol$ &   $Vol$ \\ 

	 \hline	

 ``Stimulus''   $\rightarrowtail$ $y$  & $Vol$ & $Vol$ & $Vol$ & $Vol$ &  $Vol$ & $Vol$ &   $Vol$ \\ 
     
     \hline
     
 ``Stimulus check''   $\rightarrowtail$ $y$  & - &  -& -  &  - &  - & -&  - \\ 
    \hline
   ``irs stimulus''    $\rightarrowtail$ $y$     &  & P & -  &  - &  - & $Vol$ &  $Vol$ \\ 
  
           		 \hline	
\end{tabular}
\end{table*}


\section*{References}



\bibliographystyle{elsarticle-harv}\biboptions{authoryear,comma}

\bibliography{Covid_V2}

\end{document}